\documentclass{pasj01}

\begin{document}

\title{Mass-transfer properties of overcontact systems in the Kepler eclipsing binary catalog}
\author{Shinjirou Kouzuma}%
\altaffiltext{}{Chukyo University, 101-2 Yagotohonmachi, Showa-ku, Nagoya, Aichi, 466-8666, Japan}
\email{skouzuma@lets.chukyo-u.ac.jp}


\maketitle

\begin{abstract}
We selected mass-transferring binary candidates from the catalog of Kepler eclipsing binary stars 
and investigated the dependence of the mass-transfer rate on several astrophysical quantities, 
including orbital period, semi-major axis, mass ratio, fill-out factor, temperature, and mass.
We selected these candidates using $O-C$ diagrams and calculated their mass-transfer rates. 
Primary masses were obtained from the mass--temperature relation, 
and the temperatures of the component stars were extracted from a catalog of temperatures for Kepler eclipsing binary stars.
The mass-transfer rates of overcontact systems have associations with astrophysical quantities 
that seem to differ from those of semi-detached or detached systems. 
These associations indicate that mass exchange from more- to less-massive components (from less- to more-massive components) generally becomes rapid (slow) as the mass exchange evolves. 
However, for mass exchange from more- to less-massive components, this tendency is not reasonable for binaries with a short period ($P<0.4$ d) and low mass ($M_1<1.2$ M$_\odot$) 
because the correlations of these binaries are opposite to those of binaries with long period and high mass. 
These different correlations likely arise from differences in correlation between subtypes of W UMa systems (i.e., W- and A-types). 
Alternatively, when mass exchange from more- to less-massive components reoccurs after a mass-ratio reversal, 
its properties may differ from those of the first mass exchange. 
\end{abstract}

\section{Introduction} \label{Intro}
An overcontact binary is a binary system in which both stars have exceeded their Roche lobes. 
The light curve of an overcontact system is continuously variable due to the tidally distorted shapes of the stars 
and is generally classified as belonging to the W UMa type. 
In such a system, mass transfer is likely to occur through Lagrange points. 

Mass transfer is classified into two cases: mass exchange between components and mass loss. 
In the case of mass exchange, all the mass lost by one component is gained by its companion 
and the total mass of the binary is conserved, together with the total angular momentum. 
Mass exchange leads to changes in the orbital period. 
If the rate of change of the orbital period can be obtained, the mass-exchange rate will be determined by the following equation \citep{Hilditch2001-icbs}: 
\begin{equation} \label{Conservative}
\dot{m}_1=\frac{m_1 m_2}{3(m_1-m_2)} \frac{\dot{P}}{P},
\end{equation}
where $m_1$ and $m_2$ are the masses of the two stars and $P$ and $\dot{P}$ are the orbital period and its rate of change, respectively. 
This equation indicates that the orbital period is shorter when mass exchange occurs from the more- to less-massive components 
and longer when the process occurs in the other direction. 

In the case of mass loss, the mass lost by one component escapes from the binary system. 
Mass loss is caused by phenomena such as stellar wind, outer Roche lobe overflow, or a sudden catastrophic event such as a nova or supernova. 
Assuming that mass is lost from only one component and the linear velocity of the component in its binary orbit remains constant, 
the simplest relationship between the rate of change of the period and the mass-loss rate is obtained \citep{Hilditch2001-icbs} as 
\begin{equation} \label{Non-Conservative}
\dot{m}_1=-\frac{(m_1+m_2)}{2} \frac{\dot{P}}{P}. 
\end{equation}
The orbital period must increase when mass loss occurs because $\dot{m}_1$ decreases. 

In addition to varying the shape of the Roche lobe due to mass transfer, 
the change of the orbital period alters the orbital separation. 
These changes affect the evolution of a binary star. 
For instance, Algol ($\beta$ Per) is a semi-detached binary in which the less-massive component appears to evolve earlier than the more-massive component. 
The mass-ratio reversal is explained via mass exchange between components \citep{Sarna1993-MNRAS}. 
Mass transfer also plays an important role in thermal relaxation oscillation (TRO) theory \citep{Flannery1976-ApJ,Lucy1976-ApJ,Robertson1977-MNRAS}. 
The TRO theory explains the achievement of an average thermal equilibrium in a contact system. 
According to this theory, a binary oscillates between contact and semi-detached phases via cyclic mass exchange and achieves thermal equilibrium on the average. 
As can be seen in this instance, mass transfer is associated with the evolution of a binary system. 
Thus, investigating the properties of mass transfer may solve problems associated with binary-system evolution. 

Many previous studies have focused on calculating mass-transfer rates for individual eclipsing binaries. 
However, several studies have presented the rates of change of period for some objects and showed correlations of period change with some astrophysical quantities \citep{Qian2002-MNRAS,Yang2009-AJ137}. 
Although several studies investigated the statistical properties of period change, 
few have focused on those of mass transfer; 
in other words, the statistical properties of mass transfer for binaries are not well known. 

This paper demonstrates the statistical properties of mass transfer for overcontact systems. 
Section \ref{Data} introduces the data used herein. 
In section \ref{Extraction}, we describe a method to select candidate mass-transferring binaries and to calculate mass-transfer rate. 
The dependence of this rate on astrophysical quantities is illustrated in section \ref{Dependence},  
and in section \ref{Discussion}, we discuss how the mass-transfer rate changes with the evolution of binary stars. 
Section \ref{Conclusion} summarizes our results.

\section{Data} \label{Data}
\subsection{Kepler}
The Kepler spacecraft, launched in 2009, photometrically monitored $\sim$156,000 objects within a field of $\sim$115 deg$^2$ in the direction of the constellation Cygnus. 
The main scientific goal of the Kepler Mission was to detect transits of Earth-size planets. 
To achieve this goal, Kepler must obtain a signal-to-noise ratio of 4$\sigma$ for an 84-ppm deep transit within 6.5 h. 
The overall mission design and performance were reviewed by \citet{Koch2010-ApJ}. 

The Kepler Mission offers two options for observations: long cadence (LC) and short cadence (SC). 
LC observation monitors stellar targets with a time resolution of 29.4 min, and its primary purpose is to detect transiting planets \citep{Jenkins2010-ApJ}. 
SC data, meanwhile, has a time resolution of 58.8 s and is used for applications such as asteroseismology of solar-like stars and transit-timing measurements of exoplanets \citep{Gilliland2010-ApJ}. 

Light curves derived from Kepler are also useful for studying eclipsing binaries. 
Many authors have studied such binaries using Kepler's data. 
One of the byproducts is a catalog of eclipsing-binary stars. 

\subsection{Kepler eclipsing binary catalog}
\citet{Prsa2011-AJ} presented a comprehensive catalog of eclipsing binary stars observed by Kepler (hereafter referred to as KEBC) in the first 44 days of operation. 
This catalog lists the Kepler ID, ephemeris, morphology type, physical parameters, third-light contamination levels, and principal parameters for 1,879 eclipsing binaries. 
Principal parameters (i.e., temperature ratios, photometric mass ratios, fill-out factors, and sin $i$ for overcontact systems, 
as well as the temperature ratios and sums of the fractional radii for detached and semi-detached systems) are determined via neural-network analysis of the phased light curves. 

The KEBC was updated by the second data release \citep{Slawson2011-AJ}. 
The revised catalog contains 2,165 eclipsing binaries: 1,261 detached, 152 semi-detached, 469 overcontact, 137 ellipsoidal variable, and 147 uncertain or unclassified systems. 
Principal parameters in the initial catalog are also provided in the revised version. 
In this paper, we use the revised KEBC.

\section{Principal parameters for mass-transferring binaries}\label{Extraction}
During mass transfer in a binary system, the orbital period changes.
An $O-C$ diagram is suitable for measuring the period change. 
$O-C$ values are calculated by subtracting the calculated times of the minima from those observed. 
Equations (\ref{Conservative}) and (\ref{Non-Conservative}) show that stable mass transfer causes a constant change in orbital period. 
In this situation, the $O-C$ curve should be a parabolic shape over the long term. 

This paper assumes that parabolic-period variations are solely due to mass transfer. 
Other processes that can change orbital periods are discussed in section \ref{Discussion}. 
In addition, it is assumed that mass loss does not change the specific angular-momentum per unit mass of a binary. 
In this study, we used data from Kepler's primary mission; therefore, 
candidates for mass-transferring binaries were selected on the basis of $O-C$ diagrams with time ranges between three years and a half up to four years. 

We first extracted LC photometric data for the eclipsing binaries in the KEBC and then measured the observed times of the primary minima. 
The following ephemeris determines the calculated values of these times: 
\begin{equation}
\textnormal{Min. I}=BJD_0+P \cdot E,
\end{equation}
where $BJD_0$ and $P$ are derived from the KEBC. 
We manually selected $O-C$ curves showing a parabolic shape over the long term by visual inspection. 
Using the least-squares method, we fit the $O-C$ curves with a quadratic curve described as 
\begin{equation}
(O-C)=c_2 \cdot E^2 + c_1 \cdot E + c_0, 
\end{equation}
where $c_2$, $c_1$, and $c_0$ are coefficients determined by the fitting. 
The value of $dP/dt$ is derived from the relation $dP/dE=2c_2$. 

\begin{figure}
\includegraphics[width=\hsize]{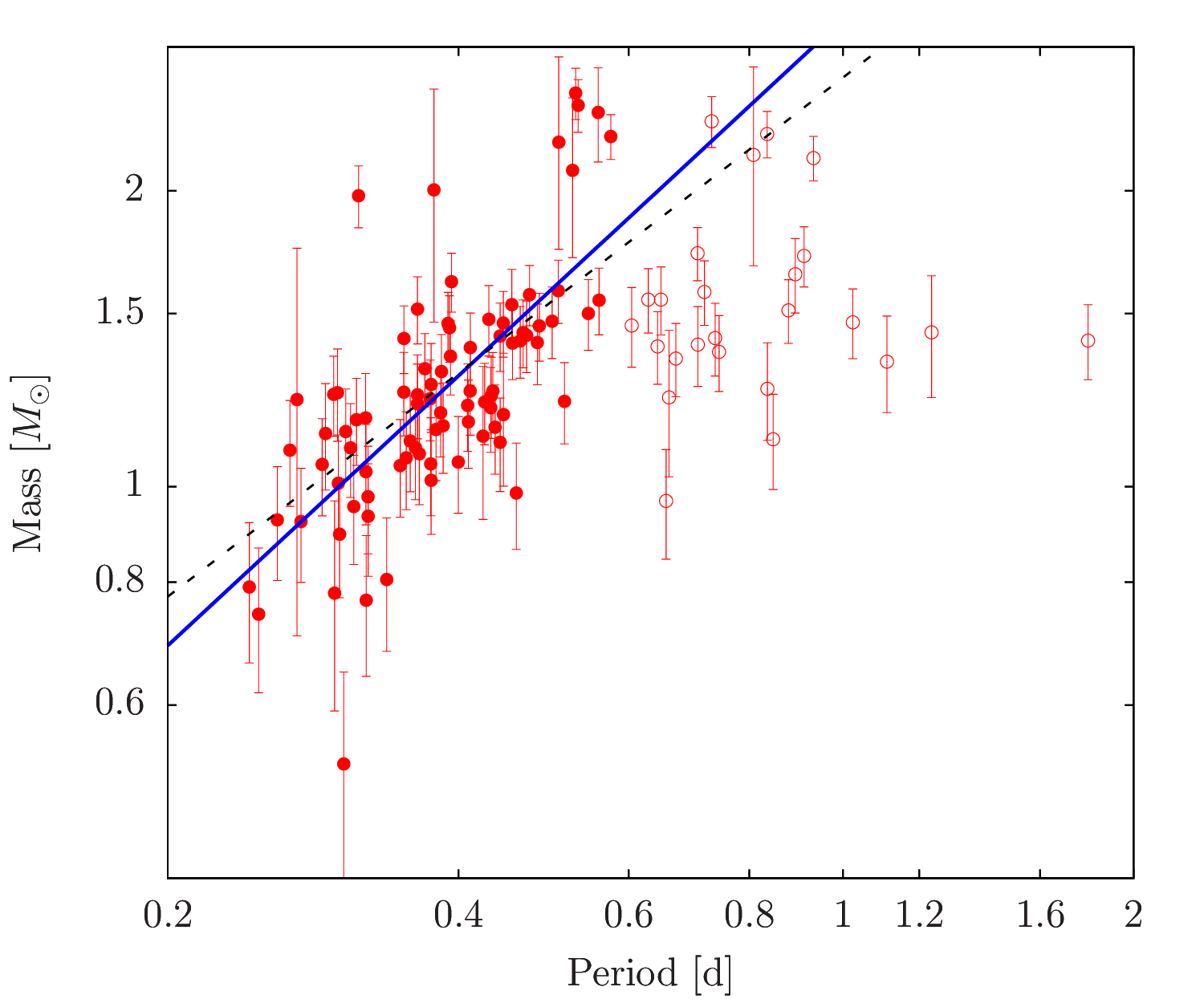}
\caption{Orbital period versus estimated mass. 
Solid and open circles represent binaries with $P<0.6$ d and $P>0.6$ d, respectively. 
The blue line shows the least-squares line for the binaries with $P<0.6$ d, 
and the dotted line represents the period--mass relation for contact binaries in \citet{Gazeas2008-MNRAS}. 
}\label{Period-Mass}
\end{figure}

The masses of both components are necessary to determine the mass transfer rate. 
KEBC includes the photometric mass ratio for overcontact systems but does not have any information about individual masses. 
If the mass of a component can be determined, the masses are computed with the mass ratio. 
In this study, we therefore calculate the masses of primary components on the basis of a mass--temperature relation in \citet{Harmanec1988-BAICz}. 
The temperature of the primary component was derived from a catalog of temperatures for Kepler eclipsing binary stars \citep{Armstrong2014-MNRAS}. 
This catalog provides primary and secondary stellar temperatures calculated with a spectral-energy distribution ranging from optical to near-infrared. 
To confirm our estimation of mass, we used a period--mass relation for contact binaries, 
which has been reported by several authors \citep{Qian2003-MNRAS,Gazeas2006-MNRAS,Eker2006-MNRAS,Gazeas2008-MNRAS}. 
\citet{Gazeas2008-MNRAS} provided power-law relations for contact binaries, and \citet{Deb2011-MNRAS} confirmed that a sample of 54 contact binaries generally obey the period--mass relation derived by \citet{Gazeas2008-MNRAS}. 
We compared our result with that reported by \citet{Gazeas2008-MNRAS}. 
Figure \ref{Period-Mass} shows the relationship between orbital period and mass, together with that for the primaries in \citet{Gazeas2008-MNRAS}. 
The distribution of binaries with $P>0.6$ d differs from that of objects with $P<0.6$ d. 
This is because most binaries with $P>0.6$ d should not be overcontact systems, as discussed in section \ref{Dependence}. 
Accordingly, we calculated the least-squares line for the binaries with $P<0.6$ d:  
\begin{equation}
\log M_1=(0.911 \pm 0.104) \log P+(0.475 \pm 0.044), 
\end{equation}
which agrees reasonably well with the correlation reported by \citet{Gazeas2008-MNRAS}. 

We calculated three types of mass-transfer rates using equations (\ref{Conservative}) and (\ref{Non-Conservative}). 
In the case of mass exchange between components, 
when $\dot{P}$ is negative, mass exchange occurs from the more-massive to less-massive components (throughout this paper we refer to this process as MEML). 
By contrast, when $\dot{P}$ is positive, mass exchange occurs from the less-massive to more-massive components (hereafter MELM). 
Mass loss (ML) occurs only when $\dot{P}$ is positive. 
The errors in the mass ratio, which are necessary to calculate those in the mass-transfer rate, are not provided in the KEBC. 
Therefore, we estimated this error on the basis of Figures 9 and 13 in \citet{Prsa2011-AJ}, together with that for the fill-out factor. 

Finally, we obtained mass-transfer rates for 111 overcontact binaries in the KEBC. 
Table \ref{Candidates} summarizes the parameters for mass-transferring binaries.

\section{Dependence of mass-transfer rates on astrophysical quantities}\label{Dependence}
\begin{figure}
\includegraphics[width=\hsize]{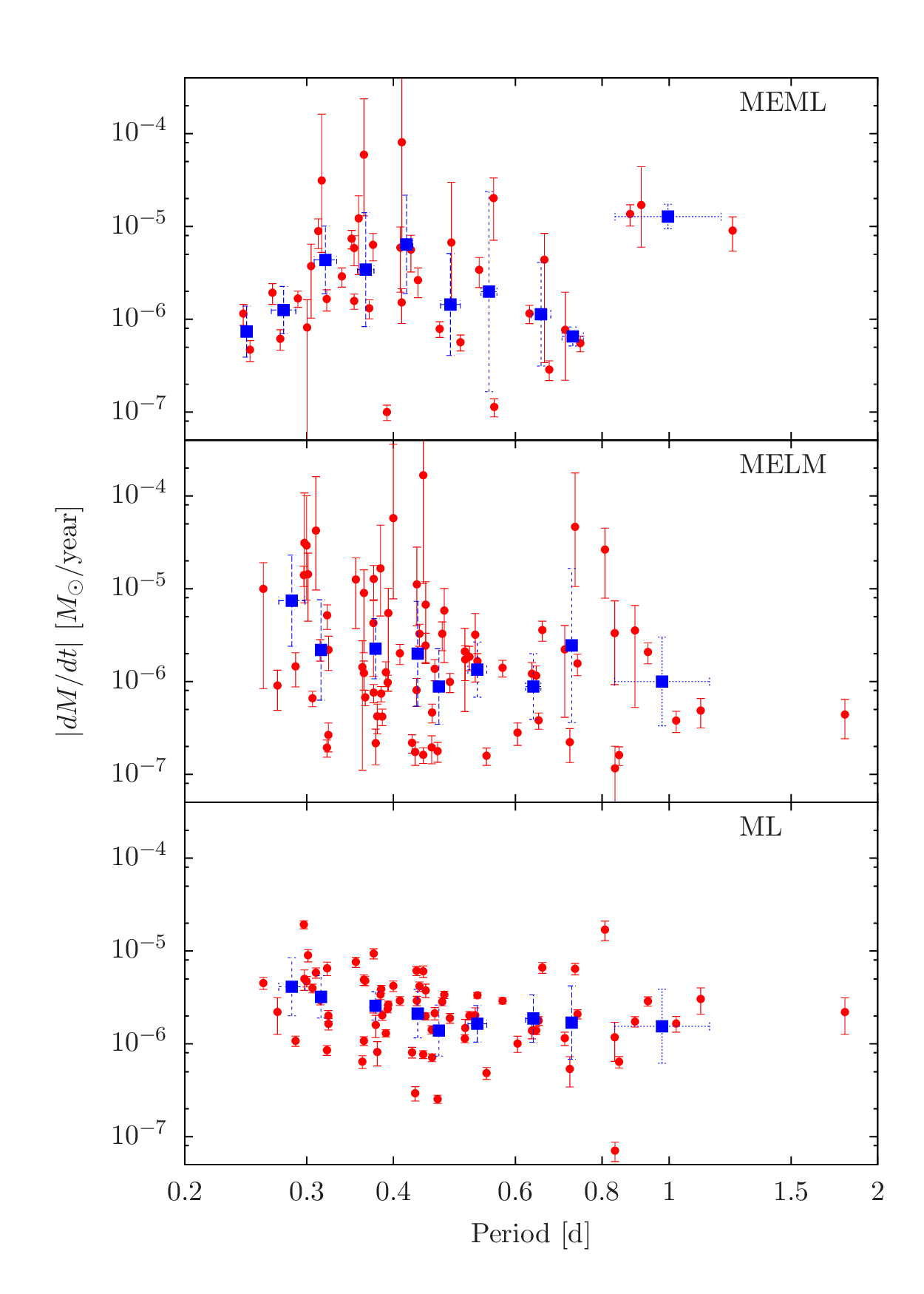}
\caption{Orbital period versus mass-transfer rate. 
The relations for MEML are shown in the top panel, those for MELM are shown in the middle panel, and the mass-loss rate is shown in the bottom panel. 
Blue solid squares represent mean values that are appropriately divided into each bin. 
}\label{per-mass_trans}
\end{figure}

\begin{figure}
\includegraphics[width=\hsize]{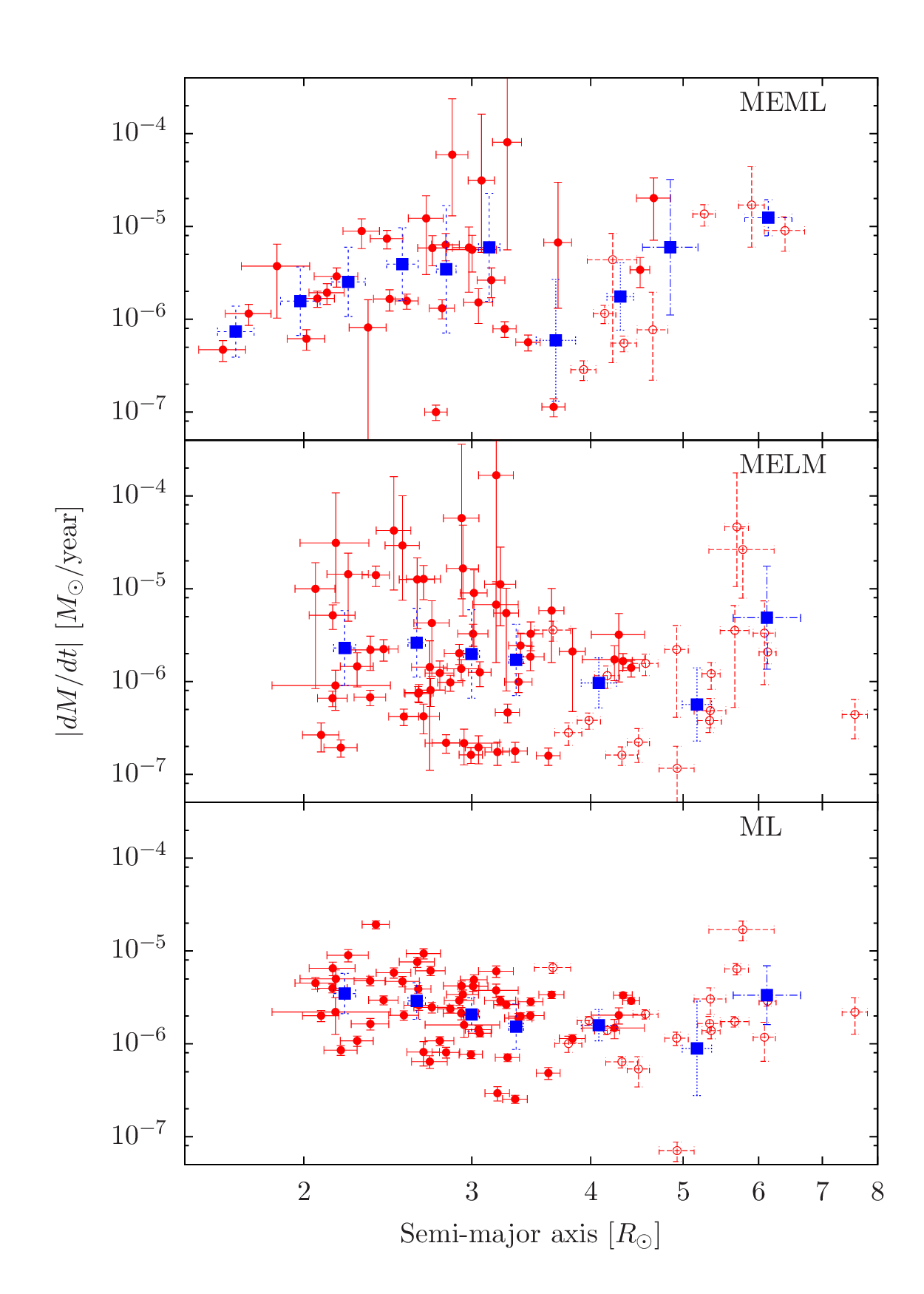}
\caption{Semi-major axis versus mass-transfer rate. 
Symbols are the same as in figures \ref{Period-Mass} and \ref{per-mass_trans}. 
}\label{a-mass_trans}
\end{figure}

\subsection{Orbital period and semi-major axis}\label{P-Semi-major}
Figure \ref{per-mass_trans} shows the relations between orbital period and mass-transfer rate. 
A general tendency for the MELM and ML samples is that the mass-transfer rates decrease with increasing orbital period within $P<0.6$ d.  
The Spearman's rank-correlation coefficients ($r$) for MELM and ML are $r=-0.226$ ($p=0.097$) and $r=-0.347$ ($p=0.010$) respectively. 
The values of $p$ are $p$-values are calculated on the basis of null hypothesis that there is no monotonic relationship between the two variables.
Although the MEML sample also has a negative correlation within $0.4$ d $<P<0.8$ d ($r=-0.562$, $p=0.024$), 
there is a positive correlation below $P\simeq0.4$ d  ($r=0.305$, $p=0.203$).
Above $P\simeq0.6$--$0.8$ d, there are no negative associations in all samples, unlike the case below this value.

Because the vast majority of contact stars have periods shorter than 0.6 d \citep{Rucinski2007-MNRAS}, 
a small fraction of binaries with periods longer than 0.6 d should be contact systems.
In addition, the classifications of binaries in the KEBC are not based on accurate modeling according to \citet{Prsa2011-AJ}. 
If most binaries with $P>0.6$ d are semi-detached or detached, 
then the properties of mass-transfer for contact differ from those for other systems. 
This agrees with a result from section \ref {Extraction}, 
namely that the period--mass relation for binaries with $P<0.6$ d differs from that for binaries with $P>0.6$ d. 
Accordingly, most binaries with $P>0.6$ d are less likely to be contact systems. 
Indeed, such an opposite correlation has already been reported by some authors.
For example, \citet{Qian2002-MNRAS} exhibited a possible correlation between $dP/dt$ and $P$ for near-contact binaries. 
They demonstrated that the rate of change of the orbital period increases as orbital period increases. 
\citet{Yang2009-AJ137} reported a correlation between orbital period and the rate of period decrease for Algol-type binaries. 
Their sample showed a correlation similar to that of \citet{Qian2002-MNRAS}. 
This positive correlation is opposite to the negative correlation of our sample objects with $P<0.6$ d. 

A relation between semi-major axis and mass-transfer rate is similar to the $P$--$\dot{M}$ relation. 
In this paper, we calculated the semi-major axis using Kepler's third law: 
\begin{equation}
\frac{a^3}{P^2}=\frac{G}{4\pi^2}(M_1+M_2). 
\end{equation}
As shown in figure \ref{a-mass_trans}, mass-transfer rates decrease with increasing semi-major axis below $a\simeq4$ R$_\odot$. 
However, the slope for the MEML sample appears to change around $a=2.5$--$3$ R$_\odot$. 
All samples seem to be positively correlated with semi-major axis above $a\simeq4$ R$_\odot$. 
Because the semi-major axis strongly depends upon the orbital period, 
it is natural that associations in $P$--$\dot{M}$ relations should be similar to those in $a$--$\dot{M}$ relations.

\begin{figure}
\includegraphics[width=\hsize]{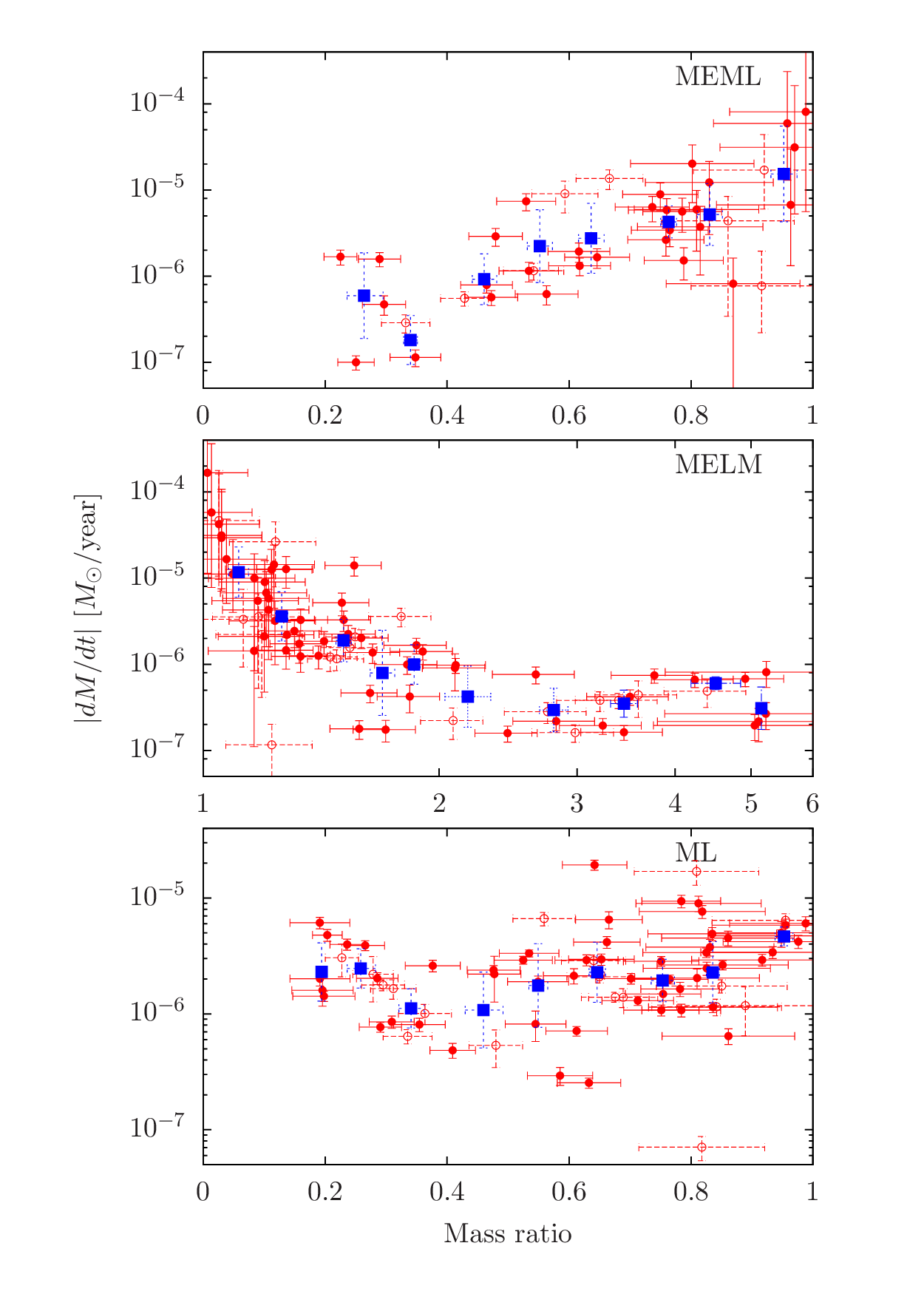}
\caption{Mass ratio versus mass-transfer rate. 
The mass ratios for ML objects are found to be between 0 and 1.
Symbols are the same as in figures \ref{Period-Mass} and \ref{per-mass_trans}.}\label{mass_ratio-mass_trans}
\end{figure}

\subsection{Mass ratio}\label{Mass_ratio}
Figure \ref{mass_ratio-mass_trans} shows the relations with the mass ratio. 
Mass ratio is defined by $q=M_{\textnormal{\scriptsize acc}}/M_{\textnormal{\scriptsize d}}$, 
where $M_{\textnormal{\scriptsize acc}}$ and $M_{\textnormal{\scriptsize d}}$ are the masses of the accretor and donor stars, respectively. 
For the ML sample, we compute the mass ratio to fall between 0 and 1.

\citet{Qian2001-MNRAS} claimed, based on their sample of W-type contact binaries, that systems showing secular-period decrease usually have a lower mass ratio ($q<0.4$) 
and periods of systems with higher mass ratios ($q>0.4$) usually show long-term increases. 
However, there are no significant differences in our samples. 

The MEML sample has a positive correlation over the whole mass-ratio range ($r=0.667$, $p<0.001$). 
By contrast, the MELM sample has a negative correlation within $q<2$ ($r=-0.694$, $p<0.001$). 
Above $q=2$--$3$, the mass-exchange rate for MELM objects is nearly constant or slightly increasing with increasing mass ratio. 
These associations indicate that mass exchange becomes more rapid when the masses of two stars are closer together. 

The distribution for the ML sample shows a shape similar to an upward parabola. 
In other words, the mass-loss rate gradually decreases below $q\simeq$ 0.5 ($r=-0.453$, $p=0.034$) and 
increases above $q\simeq$0.5 ($r=0.248$, $p=0.079$) with increasing mass ratio. 
ML tends to be more rapid for mass ratios close to 0 or 1. 

\begin{figure}
\includegraphics[width=\hsize]{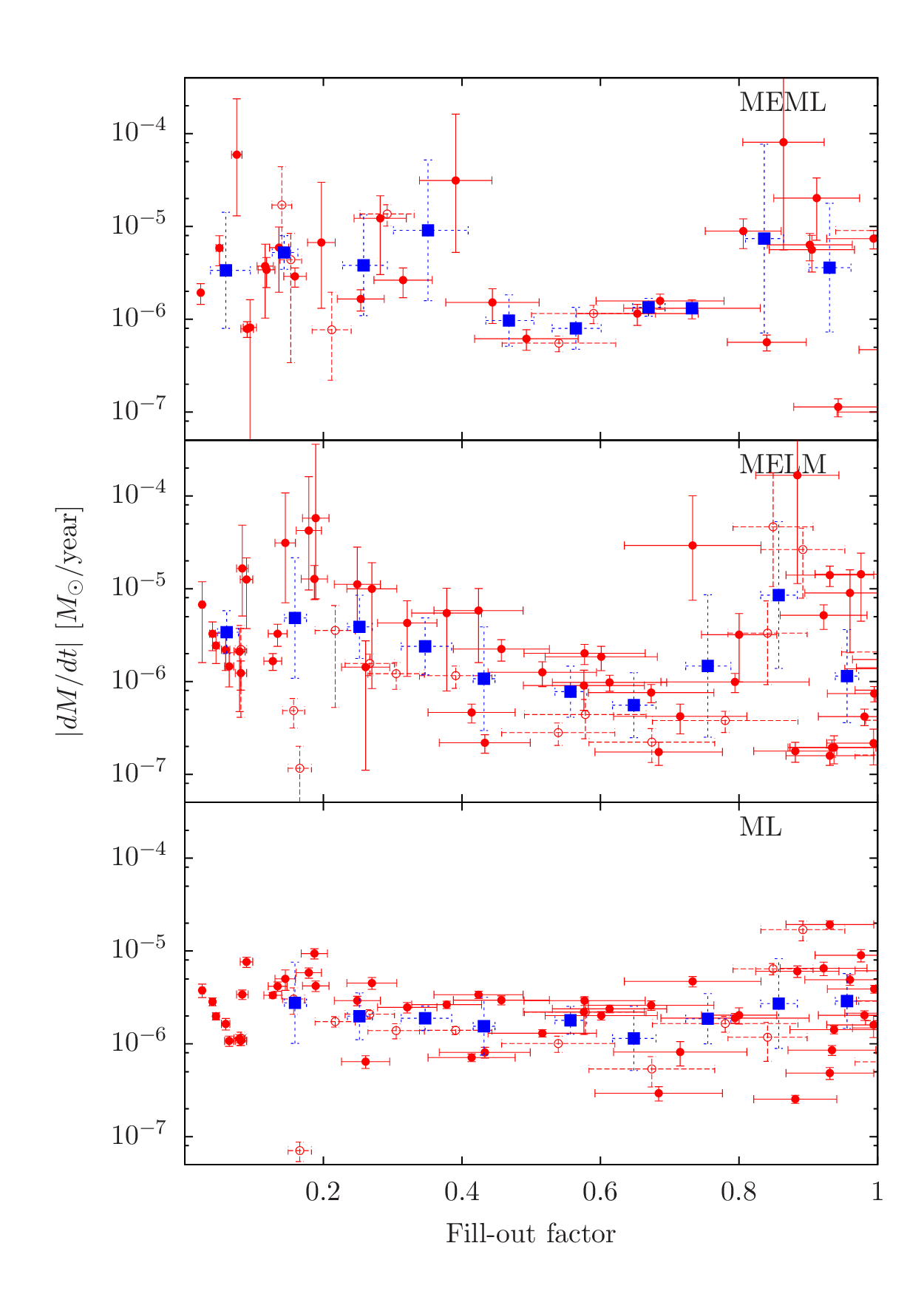}
\caption{Fill-out factor versus mass-transfer rate. 
Symbols are the same as in figures \ref{Period-Mass} and \ref{per-mass_trans}.}\label{f-mass_trans}
\end{figure}

\subsection{Fill-out factor}\label{Fillout}
Relations between fill-out factor and mass-transfer rate are shown in figure \ref{f-mass_trans}. 
The MEML and MELM samples have negative correlations below $f\simeq0.7$ ($r=-0.328$, $p=0.118$ and $r=-0.488$, $p=0.001$, respectively), 
although the correlation of the MEML sample is unclear due to low statistics in $0.4<f<0.7$. 
Furthermore, the ML sample is slightly negatively correlated. 
Above $f\simeq0.7$, there are no negative correlations and the distributions within the range are spread out. 
According to \citet{Prsa2011-AJ}, low-inclination systems with $f\sim1$ tend to be contaminated by ellipsoidal variables. 
Therefore, the scattered distributions are thought to arise from contamination. 
These associations indicate that at least within $f<0.7$,
rapid mass transfer occurs after the inner Roche lobe of a binary is filled and 
the transfer rate reduces as the outer Roche lobe is gradually filled. 

\begin{figure}
\includegraphics[width=\hsize]{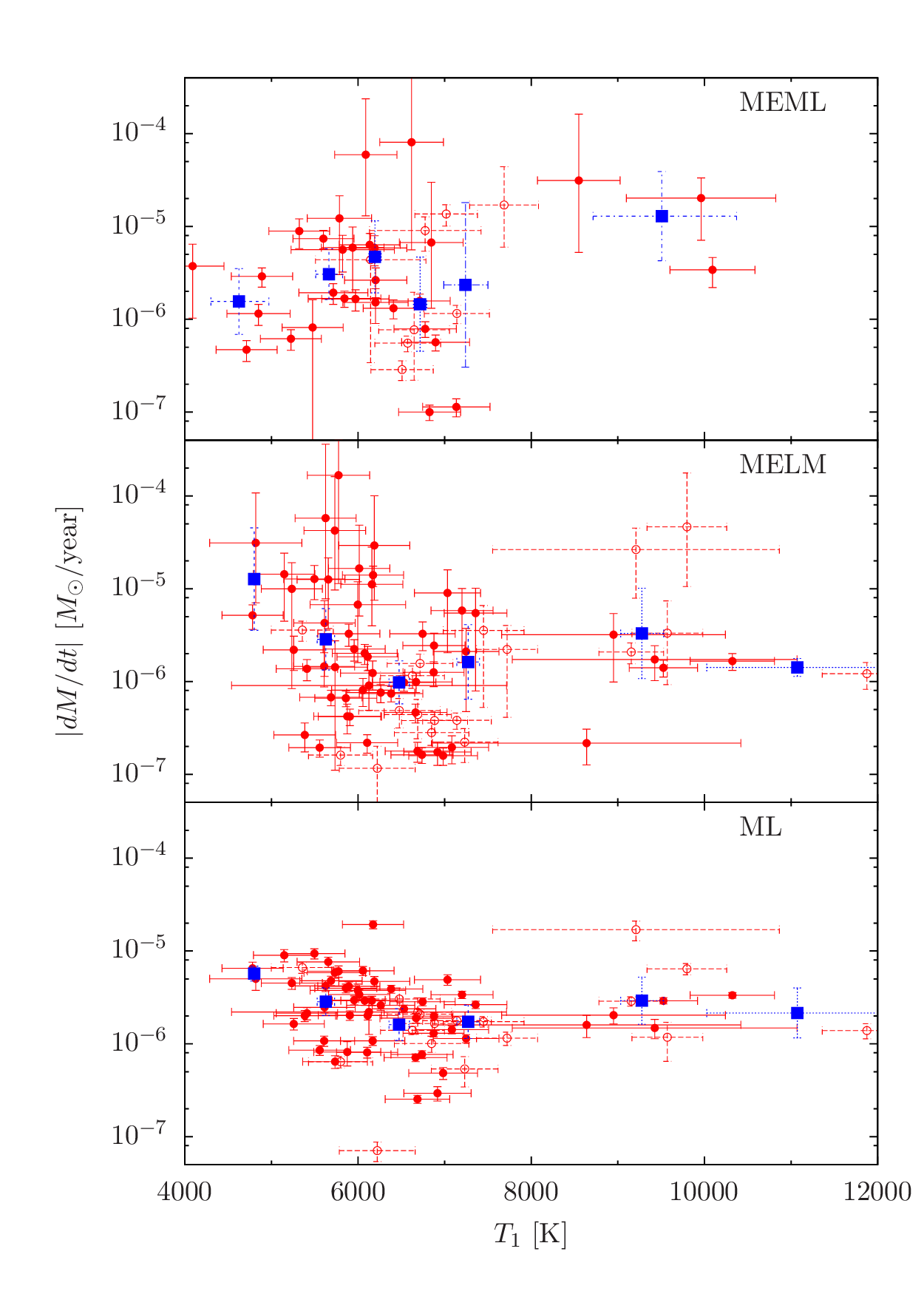}
\caption{
Temperature of primary component versus mass-transfer rate. 
Symbols are the same as in figures \ref{Period-Mass} and \ref{per-mass_trans}.\label{t1-mass_trans}
}
\end{figure}

\begin{figure}
\includegraphics[width=\hsize]{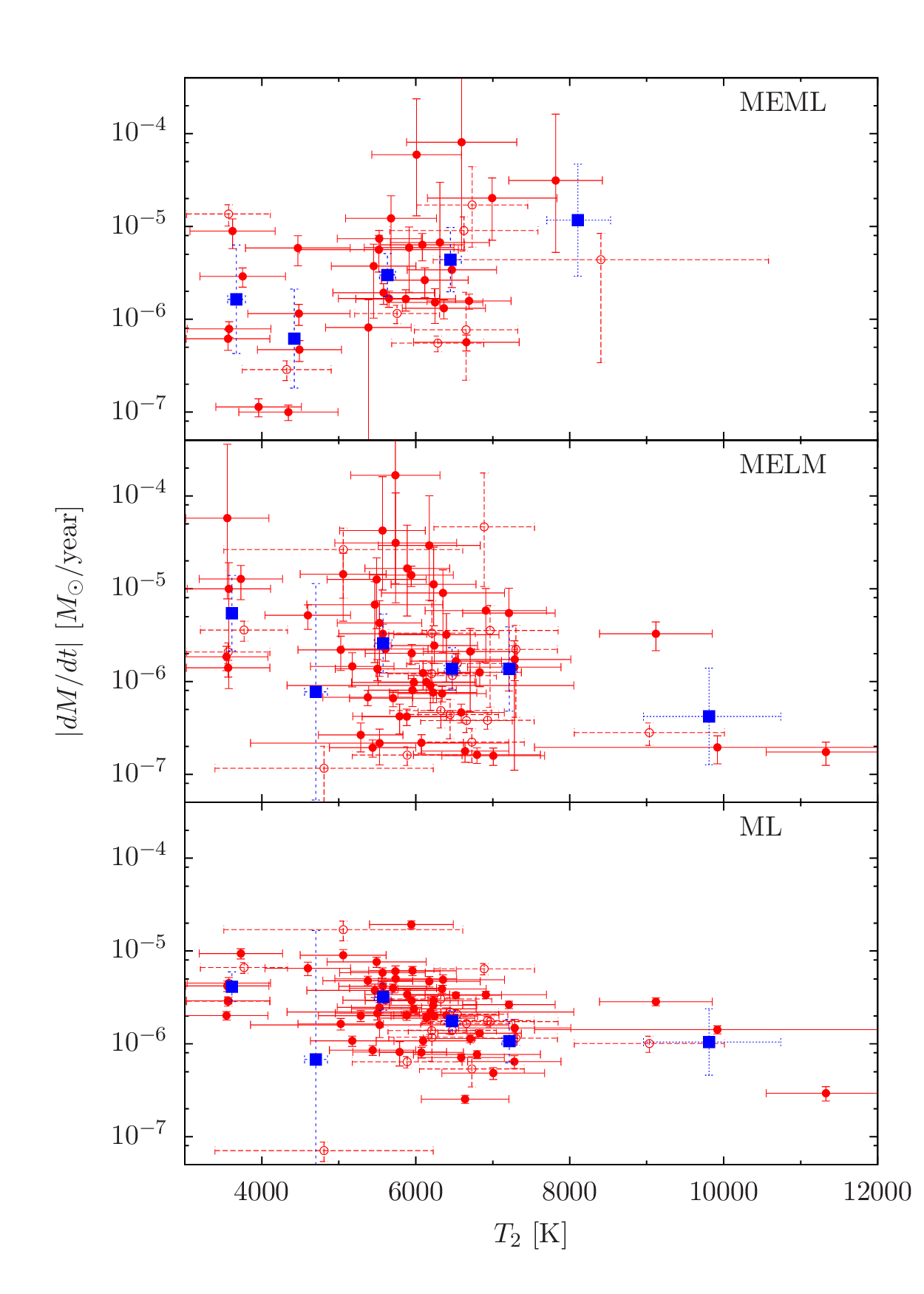}
\caption{
Temperature of secondary component versus mass-transfer rate. 
Symbols are the same as in figures \ref{Period-Mass} and \ref{per-mass_trans}.\label{t2-mass_trans}
}
\end{figure}

\subsection{Temperature}\label{Temp}
Figure \ref{t1-mass_trans} depicts the relation between the temperature of the primary component and the mass-transfer rate. 
The following associations occur below $T\simeq7000$ K: 
the MEML sample has a positive correlation below $T\simeq$6200 K ($r=0.459$, $p=0.057$) and 
a negative correlation in the range 6000 K $<T<$ 7000 K ($r=-0.328$, $p=0.198$). 
The MELM and ML samples have similar tendencies, i.e., 
the mass-transfer rate is negatively correlated with temperature ($r=-0.453$, $p=0.001$ and $r=-0.479$, $p<0.001$, respectively). 
This tendency differs from that of the MEML sample. 
Conversely, no negative associations are found above $T\simeq7000$ K for all samples. 

Characteristics differ between binaries with $P<0.6$ d and $P>0.6$ d.
Binaries with $P>0.6$ d generally have temperatures relatively higher than those of the others. 
They also have positive correlations in the scatter plots for the MELM and ML samples. 

Scatter plots for the secondary component's temperatures are shown in figure \ref{t2-mass_trans}. 
Both the MELM and ML samples show correlations similar to the $T_1$--$\dot{M}$ relationships 
($r=-0.274$, $p=0.019$ and $r=-0.436$, $p<0.001$, respectively). 
By contrast, the MEML sample seems to have a positive correlation ($r=0.336$, $p=0.039$), 
which differs from the $T_1$--$\dot{M}$ relationship. 

\begin{figure}
\includegraphics[width=\hsize]{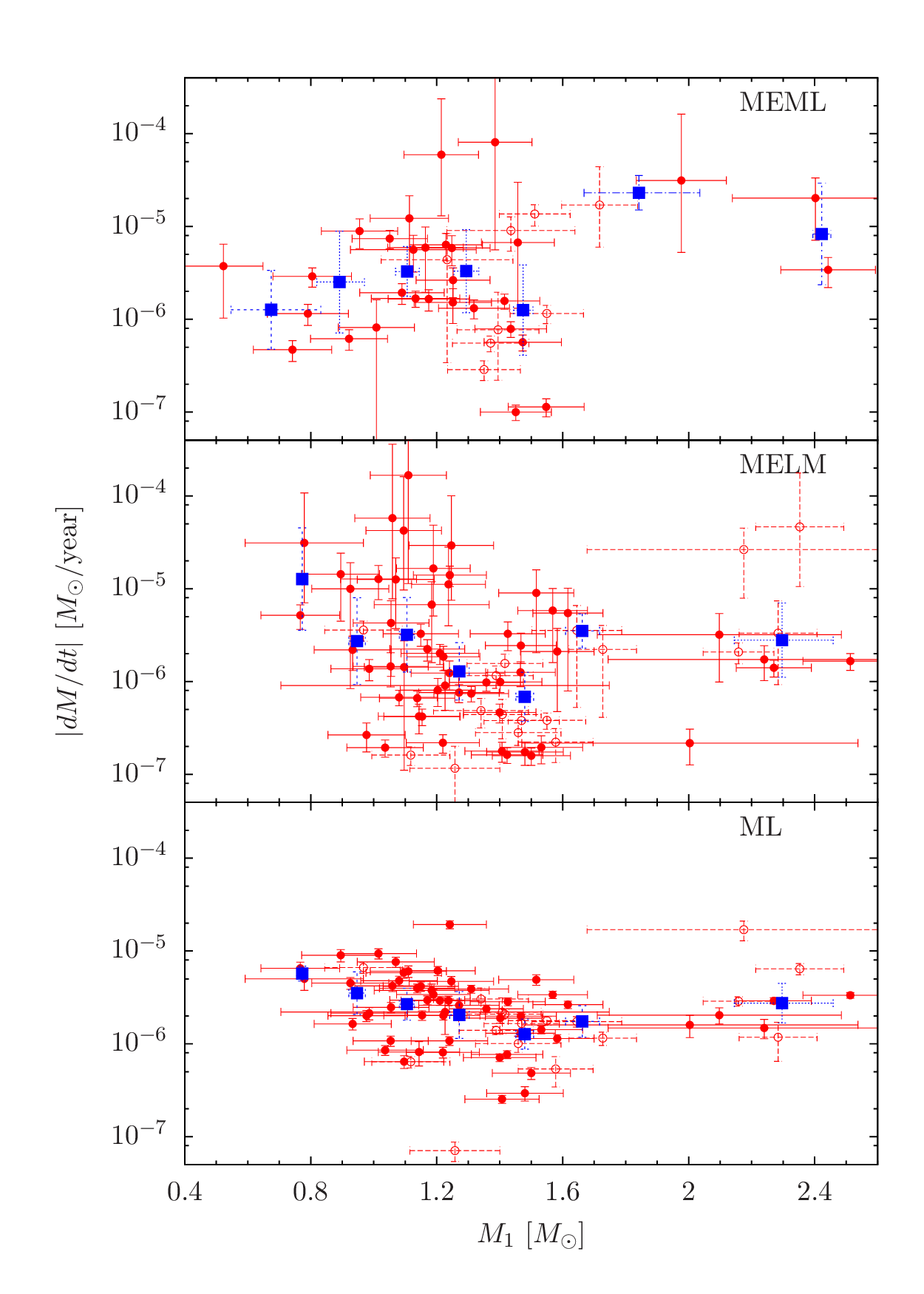}
\caption{
Primary mass versus mass-transfer rate. 
Symbols are the same as in figures \ref{Period-Mass} and \ref{per-mass_trans}.\label{m1-mass_trans}
}
\end{figure}

\begin{figure}
\includegraphics[width=\hsize]{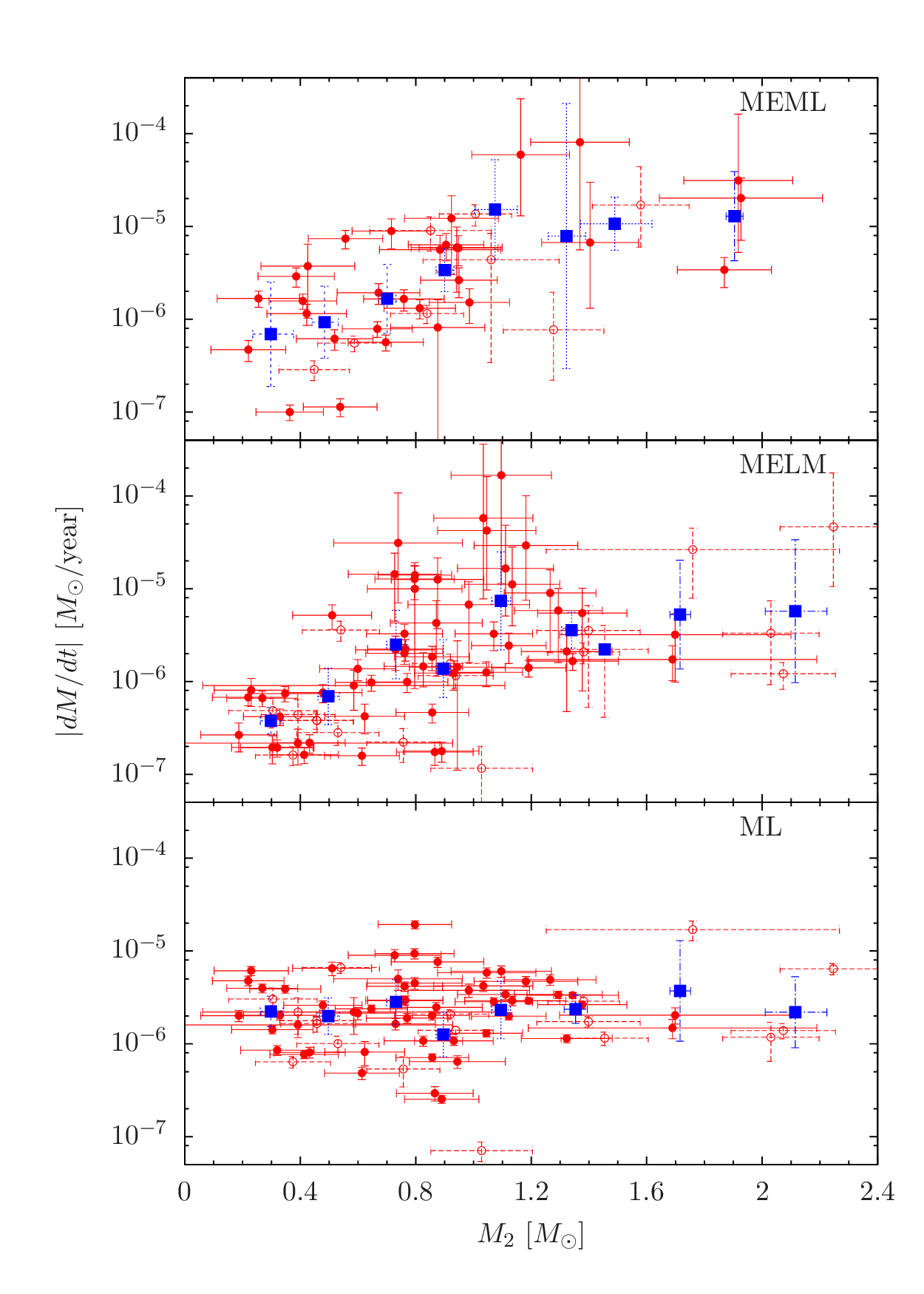}
\caption{
Secondary mass versus mass-transfer rate. 
Symbols are the same as in figures \ref{Period-Mass} and \ref{per-mass_trans}.\label{m2-mass_trans}
}
\end{figure}

\subsection{Mass}\label{Mass}
Figure \ref{m1-mass_trans} shows relations with the mass of the primary component. 
These are intrinsically the same as the $T_1$--$\dot{M}$ relationships 
because the masses are computed by the mass--temperature relation mentioned in section \ref{Extraction}. 
The association of each sample seems to change around 1.6 M$_\odot$. 
The MEML rate within $M_1<1.6$ M$_\odot$ follows a downward parabolic curve that reaches a peak around 1.2 M$_\odot$: 
$r=0.310$, $p=0.281$ for $M_1<1.2$ M$_\odot$ and $r=-0.317$, $p=0.173$ for $1.2$ M$_\odot < M_1 <1.6$ M$_\odot$. 

Relations with the mass of the secondary component are shown in figure \ref{m2-mass_trans}. 
The values of $M_2$ are calculated on the basis of the mass ratio and the mass of the primary component. 
Both the MEML and MELM rates increase as the secondary star's mass increases below $M_2\simeq1.2$ M$_\odot$ 
($r=0.588$, $p<0.001$ and $r=0.569$, $p<0.001$, respectively), 
which differ from the relationships above this mass. 
ML rates have no significant dependence on mass and assume nearly constant values.

\section{Discussion}\label{Discussion}
\subsection{Mass-transfer properties in binary systems}
Overcontact binaries are binary systems in which both components have exceeded their Roche lobes. 
Hence, mass transfer is expected to occur in such systems. 
This subsection discusses how mass transfer changes with the evolution of a binary system, 
assuming that long-term period changes are responsible only for mass transfer. 

The most plausible scenario for the formation of contact binaries is that they evolved from detached systems through angular-momentum loss 
\citep{Vilhu1982-AA,Paczynski2006-MNRAS,Stepien2006-AcA347}.
When a more massive component fills its Roche lobe, 
a detached binary system becomes semi-detached and MEML is expected to begin in the system. 
If MEML continues after filling the Roche lobe of the less-massive component, the mass-exchange process is also observed in a contact phase. 
MEML results in an increase in $q$ and $M_2$ and a decrease in $P$, $a$, and $M_1$. 
Figures \ref{per-mass_trans}, \ref{a-mass_trans}, \ref{mass_ratio-mass_trans}, \ref{m1-mass_trans}, and \ref{m2-mass_trans} demonstrate that 
the correlations between the MEML rate and these parameters show the same trend. 
In other words, the MEML rate of a contact system increases as the binary evolves. 
This relationship indicates that the fill-out factor should decrease with evolution (figure \ref{f-mass_trans}). 
Notably, this is reasonable for binaries with $P>0.4$ d and $M_1>1.2$ M$_\odot$ 
because downward parabolic associations appear in the $P$-- and $M_1$--$\dot{M}$ relations. 
However, the MEML rate of a binary with relatively short orbital period (i.e., $P<0.4$ d) may decrease as the mass-transfer evolves. 

MELM will occur if the mass exchange from initially more- to initially less-massive components continues after a mass-ratio reversal. 
This scenario has been considered by \citet{Stepien2006-AcA199}. 
Alternatively, MELM probably occurs in the cycle predicted by the TRO theory \citep{Flannery1976-ApJ,Lucy1976-ApJ,Robertson1977-MNRAS}. 
MELM results in a decrease in $M_2$ and an increase in $P$, $a$, $q$, and $M_1$. 
In this situation, the MELM rate decreases with the evolution of the binary and the fill-out factor should increase. 

Another notable feature is that there are parabolic associations in the relationships with orbital period, semi-major axis, primary temperature, and primary mass. 
The parabolic associations indicate that a correlation changes to the opposite of the correlation around a value $P\simeq0.4$ d, $a\simeq3$ R$_\odot$, $T_1\simeq6000$ K, and $M_1\simeq1.2$ M$_\odot$. 
W UMa binaries are subdivided into two groups by \citet{Binnendijk1970-VA}: W- and A-types. 
These subtypes are distinguishable by orbital period and spectral type. 
In general, W-type systems have orbital periods shorter than 0.5 d and G--K spectra, 
whereas A-type systems have orbital periods longer than 0.3 d and A--F spectra \citep{Webbink2003-ASPC,Gazeas2006-MNRAS}. 
Thus, the two subtypes can be roughly separated by the values of $P\simeq0.4$ d and $T\simeq6000$ K. 
Accordingly, the parabolic associations are probably caused by differences in correlation between the subtypes. 
Alternatively, these associations may appear 
because MEML reoccurs after a mass-ratio reversal and 
the dependence of the MEML rate differs from that for the original mass transfer. 

The $O-C$ diagrams of our sample binaries have a time range of about four years at most. 
The rate of orbital-period change computed from an $O-C$ diagram with a short time span tends to be large. 
In practice, while all of the binaries in table \ref{Candidates} have mass-transfer rates higher than $10^{-7}$ M$_\odot$ yr$^{-1}$, 
binaries with rates as low as  $10^{-8}$ M$_\odot$ yr$^{-1}$ have been reported (e.g., \cite{Zhu2009-AJ} and references therein). 
The properties of binaries with lower and higher mass-transfer rates do not need to be in agreement with each other. 
In other words, binaries with lower mass-transfer rates may exhibit properties different from those with higher ones. 
Accordingly, the properties of binaries with low mass-transfer rates should also be investigated with other data. 

\subsection{Other possible processes}
Mass transfer is a plausible process for explaining the parabolic $O-C$ curves of overcontact binaries 
because the components of an overcontact binary exceed their Roche lobes. 
However, periodic (particularly sinusoidal-like) oscillation in $O-C$ value can cause confusion with the parabolic curve in an $O-C$ diagram. 
Such confusion tends to arise when the time range of an $O-C$ diagram is short. 
Periodic oscillations in orbital period have been explained not by mass transfer but by cyclic magnetic activity or the presence of a third body. 
Hence, confusion between parabolic and periodic curves may make our results unreliable. 
We consider two possible processes. 

\citet{Applegate1992-ApJ} demonstrated that orbital period modulations can be explained via the gravitational coupling of the orbit 
to variations in the shape of a magnetically active star in the system. 
Binary systems with at least one convective star were suggested to have orbital-period changes due to this mechanism. 
As discussed in the previous subsection, relationships between temperature and mass transfer show parabolic associations with vertices at $T\sim 6000$ K. 
Stars with $T\lesssim6000$ K generally have convective envelopes and are magnetically active due to stellar dynamos. 
Therefore, some binaries in which a component has $T\lesssim6000$ K may possibly change their orbital periods through this mechanism. 
Confusion between mass-transfer and this mechanism might contribute to changes in correlation at $T\simeq6000$ K or 
to dispersed distributions. 

Another possible process is the light-travel time effect (LTTE) related to the motion around a third body \citep{Irwin1959-AJ}. 
Many authors have explained that cyclic period modulations are due to the LTTE (e.g., \cite{Frieboes1973-AAS,Borkovits1996-AAS}). 
Moreover, a third companion of several binaries has been confirmed via adaptive-optics observations \citep{Tokovinin2006-AA,Rucinski2007-AJ}. 
If sinusoidal-like period change caused by the LTTE does not appear as a periodic curve in an $O-C$ diagram, 
it is difficult to distinguish between parabolic and periodic curves using this diagram alone. 
However, most $O-C$ curves for our sample binaries show superposition of cyclic modulation upon a parabolic curve. 
Such superposition, which often appears, has been interpreted as both mass-transfer and the LTTE (e.g., \cite{Yang2011-PASP} and references therein). 
Furthermore, if the period changes of a majority of binaries are due to LTTE, 
there should be no correlation between binary parameters and mass-transfer rates 
because if the period-change rate or parameters related to this rate are correlated with binary parameters, 
the correlations suggest that third bodies affect the physics of binary systems.
Alternatively, they suggest that a third body is formed under the influence of the binary star.
Therefore, although some of our sample binaries might show orbital-period change due to only LTTE rather than mass transfer, 
they should be a minority. 
Instead, confusion between mass transfer and the LTTE should contribute to dispersed distributions.

\section{Summary and Conclusions}\label{Conclusion}
We have investigated the statistical properties of mass transfer for overcontact binaries in the KEBC, 
assuming that the simplest mass transfer occurs in binary systems. 
We have shown that the mass-transfer rate is associated with the astrophysical quantities of binary systems. 
Moreover, associations differ between the MEML and MELM samples, 
although the ML sample has dependence similar to that of the MELM sample. 

Most binaries with $P>0.6$ d are likely to be contaminated by semi-detached or detached systems. 
Furthermore, their properties differ from those of binaries with $P<0.6$ d 
in terms of relation to orbital period, semi-major axis, primary temperature, and primary mass. 
We inferred that the difference in the mass-transfer properties is due to the difference between types of binary systems. 
To confirm this, the properties for semi-detached and detached systems should be examined. 
Moreover, it should be confirmed that the properties of such systems are exactly different from those of overcontact systems. 

We discussed how mass-transfer rates change with the evolution of contact-binary systems and concluded the following. 
Mass exchange from more- to less-massive components becomes rapid as a binary evolves. 
By contrast, the rate of mass exchange from less- to more-massive components decreases with evolution. 
However, the properties of MEML binaries with short period ($P<0.4$ d) or low mass ($M<1.2$ M$_\odot$) differ from those with longer periods or higher mass. 
This is likely to arise from the mass-transfer properties of W-type binaries differing from those of A-type ones 
or because the properties of a MEML occurring after a mass-ratio reversal differed from those occurring before. 
We note that in practice, both processes of mass exchange between the components and mass-loss should simultaneously occur in a binary system. 
Furthermore, in magnetically active binaries, magnetized wind may have a long-term effect on period change. 
In particular, in the case of ML, angular-momentum loss via magnetized wind shortens the orbital period, 
which is the opposite of the period change due to ML. 
If this process strongly contributes to long-term period change, the estimated mass-transfer rate will deteriorate. 

Other probable processes that may have caused orbital-period oscillations have also been discussed 
because the parabolic $O-C$ curves may be confused with periodic ones. 
Some sample binaries for which a component has $T<6000$ K might change their orbital period by the Applegate mechanism. 
These binaries are likely to affect the correlation between temperature and mass-transfer rate below $T\simeq 6000$ K. 
Although some binaries possibly show orbital-period change by only LTTE rather than mass transfer, 
we have determined them to be in the minority. 
In addition, confusion between mass transfer and other processes possibly lead to dispersed distributions. 

The masses and mass ratios for sample binaries, which are necessary to calculate mass-transfer rates, are determined by photometric 
rather than spectroscopic data. 
Accordingly, these quantities and mass-transfer rates may have large uncertainties. 
If sample binaries with spectroscopically determined absolute parameters are used, then more reliable results would be derived.

\begin{longtable}{*{10}{r}}
\caption{Principal parameters for mass-transferring overcontact binaries. \label{Candidates}}
\hline
KIC & $P$ & $a$ & $T_1$ & $f$ & $M_1$ & $M_{\textnormal{\scriptsize acc}}/M_{\textnormal{\scriptsize d}}$ & $dP/dt$ & ME rate & ML rate \\
 & days & R$_\odot$ & K &  & M$_\odot$ &  & $\times 10^{-7}$ d yr$^{-1}$ & $\times 10^{-7}$ M$_\odot$ yr$^{-1}$  & $\times 10^{-7}$ M$_\odot$ yr$^{-1}$ \\
\hline
\endfirsthead
\hline
KIC & $P$ & $a$ & $T_1$ & $f$ & $M_1$ & $M_{\textnormal{\scriptsize acc}}/M_{\textnormal{\scriptsize d}}$ & $dP/dt$ & ME rate & ML rate \\
 & days & R$_\odot$ & K &  & M$_\odot$ &  & $\times 10^{-7}$ d yr$^{-1}$ & $\times 10^{-7}$ M$_\odot$ yr$^{-1}$  & $\times 10^{-7}$ M$_\odot$ yr$^{-1}$ \\
\hline
\endhead
\hline
\endfoot
\hline
\endlastfoot
\hline
 2159783 & $ 0.37388$ & $ 2.82$ & $ 6135$ & $ 0.902$ & $ 1.23$ & $0.736$ & $ -20.707 \pm    0.398$ & $ -63.386$ &     ---  \\ 
 2437038 & $ 0.26768$ & $ 2.11$ & $ 5715$ & $ 0.023$ & $ 1.09$ & $0.616$ & $  -8.888 \pm    0.391$ & $ -19.343$ &     ---  \\ 
 2444187 & $ 0.39016$ & $ 3.06$ & $ 6873$ & $ 0.516$ & $ 1.47$ & $1.403$ & $   4.045 \pm    0.150$ & $  12.572$ & $  13.016$  \\ 
 2715007 & $ 0.29711$ & $ 2.38$ & $ 6173$ & $ 0.931$ & $ 1.24$ & $1.558$ & $  56.213 \pm    0.263$ & $ 140.328$ & $ 192.843$  \\ 
 2854432 & $ 0.32233$ & $ 2.09$ & $ 5385$ & $ 1.090$ & $ 0.98$ & $5.225$ & $  11.135 \pm    0.342$ & $   2.663$ & $  20.105$  \\ 
 2854752 & $ 0.47043$ & $ 3.46$ & $ 6745$ & $ 0.040$ & $ 1.43$ & $1.332$ & $  10.737 \pm    0.406$ & $  32.711$ & $  28.486$  \\ 
 2858322 & $ 0.43640$ & $ 3.01$ & $ 5894$ & $ 0.134$ & $ 1.15$ & $1.511$ & $  19.024 \pm    0.729$ & $  32.714$ & $  41.640$  \\ 
 3104113 & $ 0.84679$ & $ 4.31$ & $ 5799$ & $ 1.038$ & $ 1.12$ & $2.983$ & $   7.266 \pm    0.595$ & $   1.612$ & $   6.403$  \\ 
 3127873 & $ 0.67153$ & $ 3.93$ & $ 6508$ & $ 1.100$ & $ 1.35$ & $0.332$ & $  -8.635 \pm    1.146$ & $  -2.876$ &     ---  \\ 
 3221207 & $ 0.47383$ & $ 3.64$ & $ 7201$ & $ 0.424$ & $ 1.57$ & $1.212$ & $  11.203 \pm    0.253$ & $  58.291$ & $  33.831$  \\ 
 3745184 & $ 0.30423$ & $ 1.87$ & $ 4091$ & $ 0.116$ & $ 0.52$ & $0.815$ & $ -14.815 \pm    0.208$ & $ -37.374$ &     ---  \\ 
 3756730 & $ 0.37916$ & $ 2.67$ & $ 5876$ & $ 0.715$ & $ 1.14$ & $1.835$ & $   3.510 \pm    0.943$ & $   4.226$ & $   8.176$  \\ 
 3848042 & $ 0.41145$ & $ 3.27$ & $ 6618$ & $ 0.864$ & $ 1.39$ & $0.988$ & $  -8.636 \pm    0.597$ & $-807.609$ &     ---  \\ 
 3936357 & $ 0.36915$ & $ 2.79$ & $ 6409$ & $ 0.732$ & $ 1.32$ & $0.617$ & $  -6.850 \pm    0.137$ & $ -13.151$ &     ---  \\ 
 4074532 & $ 0.35315$ & $ 2.63$ & $ 5657$ & $ 0.089$ & $ 1.07$ & $1.222$ & $  27.703 \pm    0.103$ & $ 126.014$ & $  76.302$  \\ 
 4464999 & $ 0.43416$ & $ 3.14$ & $ 6204$ & $ 0.315$ & $ 1.25$ & $0.759$ & $  -8.753 \pm    0.149$ & $ -26.454$ &     ---  \\ 
 4563150 & $ 0.27472$ & $ 2.01$ & $ 5225$ & $ 0.493$ & $ 0.92$ & $0.563$ & $  -4.287 \pm    0.190$ & $  -6.179$ &     ---  \\ 
 4569923 & $ 0.31358$ & $ 2.42$ & $ 5956$ & $ 0.457$ & $ 1.17$ & $1.531$ & $   9.575 \pm    0.300$ & $  22.412$ & $  29.530$  \\ 
 4991959 & $ 0.36094$ & $ 2.71$ & $ 5735$ & $ 0.261$ & $ 1.10$ & $1.161$ & $   2.276 \pm    0.202$ & $  14.311$ & $   6.433$  \\ 
 5015926 & $ 0.36269$ & $ 2.78$ & $ 6169$ & $ 0.081$ & $ 1.24$ & $1.331$ & $   3.592 \pm    0.109$ & $  12.370$ & $  10.756$  \\ 
 5022573 & $ 0.44172$ & $ 3.18$ & $ 5775$ & $ 0.884$ & $ 1.11$ & $1.012$ & $  24.207 \pm    1.490$ & $1668.919$ & $  60.446$  \\ 
 5123176 & $ 0.70784$ & $ 4.65$ & $ 6648$ & $ 0.212$ & $ 1.39$ & $0.916$ & $  -1.080 \pm    0.346$ & $  -7.712$ &     ---  \\ 
 5198934 & $ 0.83474$ & $ 6.09$ & $ 9571$ & $ 0.841$ & $ 2.28$ & $1.125$ & $   4.560 \pm    2.024$ & $  33.303$ & $  11.782$  \\ 
 5201619 & $ 0.50728$ & $ 3.83$ & $ 7248$ & $ 0.079$ & $ 1.58$ & $1.196$ & $   3.983 \pm    0.139$ & $  21.112$ & $  11.406$  \\ 
 5296877 & $ 0.37726$ & $ 2.95$ & $ 8639$ & $ 0.994$ & $ 2.00$ & $5.114$ & $   5.038 \pm    0.131$ & $   2.168$ & $  15.996$  \\ 
 5310387 & $ 0.44167$ & $ 2.99$ & $ 6738$ & $ 1.086$ & $ 1.42$ & $3.443$ & $   3.696 \pm    0.179$ & $   1.625$ & $   7.685$  \\ 
 5353374 & $ 0.39332$ & $ 3.26$ & $ 7357$ & $ 0.378$ & $ 1.62$ & $1.174$ & $   6.942 \pm    0.121$ & $  54.642$ & $  26.409$  \\ 
 5440746 & $ 0.48265$ & $ 3.36$ & $ 6670$ & $ 0.794$ & $ 1.40$ & $1.821$ & $   8.419 \pm    0.575$ & $   9.928$ & $  18.940$  \\ 
 5450322 & $ 0.42402$ & $ 3.00$ & $ 5823$ & $ 0.905$ & $ 1.13$ & $0.786$ & $ -17.352 \pm    0.933$ & $ -56.251$ &     ---  \\ 
 5535061 & $ 0.46343$ & $ 3.33$ & $ 6687$ & $ 0.881$ & $ 1.41$ & $1.581$ & $   1.024 \pm    0.046$ & $   1.783$ & $   2.537$  \\ 
 5703230 & $ 0.73147$ & $ 5.69$ & $ 9798$ & $ 0.849$ & $ 2.35$ & $1.047$ & $  20.433 \pm    2.248$ & $ 464.896$ & $  64.240$  \\ 
 5770431 & $ 0.39244$ & $ 2.85$ & $ 6531$ & $ 0.613$ & $ 1.36$ & $2.100$ & $   9.336 \pm    0.199$ & $   9.785$ & $  23.835$  \\ 
 5770860 & $ 0.73756$ & $ 4.57$ & $ 6716$ & $ 0.267$ & $ 1.42$ & $1.539$ & $  13.194 \pm    0.916$ & $  15.669$ & $  20.900$  \\ 
 5790912 & $ 0.38332$ & $ 2.94$ & $ 6014$ & $ 0.083$ & $ 1.19$ & $1.071$ & $  11.326 \pm    0.192$ & $ 165.764$ & $  33.981$  \\ 
 5820209 & $ 0.65609$ & $ 3.65$ & $ 5357$ & $-0.024$ & $ 0.97$ & $1.789$ & $  57.680 \pm    0.933$ & $  35.934$ & $  66.293$  \\ 
 5881838 & $ 0.30033$ & $ 2.34$ & $ 5476$ & $ 0.094$ & $ 1.01$ & $0.869$ & $  -1.099 \pm    0.208$ & $  -8.156$ &     ---  \\ 
 5951553 & $ 0.43197$ & $ 2.72$ & $ 6056$ & $ 1.038$ & $ 1.20$ & $5.231$ & $  36.899 \pm    1.347$ & $   8.097$ & $  61.211$  \\ 
 6044543 & $ 0.53209$ & $ 4.51$ & $10093$ & $ 0.118$ & $ 2.44$ & $0.765$ & $  -6.843 \pm    0.132$ & $ -34.111$ &     ---  \\ 
 6061139 & $ 0.32244$ & $ 2.35$ & $ 5258$ & $ 0.059$ & $ 0.93$ & $1.279$ & $   6.362 \pm    0.236$ & $  22.017$ & $  16.406$  \\ 
 6118779 & $ 0.36425$ & $ 2.35$ & $ 5688$ & $ 1.092$ & $ 1.08$ & $4.917$ & $  26.834 \pm    0.638$ & $   6.773$ & $  47.887$  \\ 
 6370361 & $ 0.45491$ & $ 3.27$ & $ 6664$ & $ 0.414$ & $ 1.40$ & $1.633$ & $   2.867 \pm    0.104$ & $   4.645$ & $   7.111$  \\ 
 6370665 & $ 0.93232$ & $ 6.13$ & $ 9155$ & $ 1.017$ & $ 2.16$ & $1.562$ & $  15.163 \pm    1.512$ & $  20.820$ & $  28.788$  \\ 
 6424124 & $ 0.38553$ & $ 2.55$ & $ 5906$ & $ 0.981$ & $ 1.15$ & $3.504$ & $  10.544 \pm    0.477$ & $   4.200$ & $  20.275$  \\ 
 6467389 & $ 0.28898$ & $ 2.27$ & $ 5609$ & $ 0.064$ & $ 1.05$ & $1.276$ & $   3.308 \pm    0.139$ & $  14.591$ & $  10.758$  \\ 
 6677225 & $ 0.52503$ & $ 4.28$ & $ 8951$ & $ 0.800$ & $ 2.10$ & $1.235$ & $   5.636 \pm    0.223$ & $  31.988$ & $  20.372$  \\ 
 6791604 & $ 0.52881$ & $ 4.32$ & $10322$ & $ 0.127$ & $ 2.51$ & $1.871$ & $   9.141 \pm    0.101$ & $  16.626$ & $  33.332$  \\ 
 6803335 & $ 1.11085$ & $ 5.34$ & $ 6478$ & $ 0.157$ & $ 1.34$ & $4.398$ & $  41.130 \pm   12.125$ & $   4.869$ & $  30.458$  \\ 
 7130044 & $ 0.29767$ & $ 2.16$ & $ 4819$ & $ 0.145$ & $ 0.78$ & $1.055$ & $  19.640 \pm    0.211$ & $ 312.512$ & $  50.089$  \\ 
 7339345 & $ 0.25966$ & $ 2.06$ & $ 5236$ & $ 0.270$ & $ 0.93$ & $1.162$ & $  13.620 \pm    0.139$ & $  99.886$ & $  45.161$  \\ 
 7458285 & $ 0.66063$ & $ 4.22$ & $ 6146$ & $ 0.153$ & $ 1.23$ & $0.861$ & $ -11.412 \pm    0.147$ & $ -43.818$ &     ---  \\ 
 7501230 & $ 0.89275$ & $ 5.67$ & $ 7450$ & $ 0.217$ & $ 1.64$ & $1.176$ & $  10.197 \pm    0.788$ & $  35.578$ & $  17.379$  \\ 
 7506164 & $ 0.55801$ & $ 4.66$ & $ 9961$ & $ 0.912$ & $ 2.40$ & $0.802$ & $ -34.829 \pm    0.838$ & $-202.352$ &     ---  \\ 
 7518816 & $ 0.46658$ & $ 3.25$ & $ 6775$ & $ 0.090$ & $ 1.44$ & $0.465$ & $  -8.870 \pm    0.416$ & $  -7.893$ &     ---  \\ 
 7584739 & $ 0.91156$ & $ 5.90$ & $ 7687$ & $ 0.140$ & $ 1.72$ & $0.920$ & $ -23.610 \pm    0.596$ & $-170.388$ &     ---  \\ 
 7709086 & $ 0.40947$ & $ 2.98$ & $ 5938$ & $ 0.136$ & $ 1.16$ & $0.809$ & $ -14.763 \pm    0.438$ & $ -59.257$ &     ---  \\ 
 7766185 & $ 0.83546$ & $ 4.93$ & $ 6222$ & $ 0.166$ & $ 1.26$ & $1.223$ & $   0.517 \pm    0.103$ & $   1.164$ & $   0.707$  \\ 
 7816201 & $ 0.57496$ & $ 4.41$ & $ 9526$ & $ 1.040$ & $ 2.27$ & $1.906$ & $   9.676 \pm    0.447$ & $  14.056$ & $  29.125$  \\ 
 7871200 & $ 0.24290$ & $ 1.75$ & $ 4851$ & $ 0.653$ & $ 0.79$ & $0.534$ & $  -9.270 \pm    0.161$ & $ -11.522$ &     ---  \\ 
 7878402 & $ 0.37435$ & $ 2.72$ & $ 5613$ & $ 0.321$ & $ 1.05$ & $1.211$ & $   9.605 \pm    0.247$ & $  42.762$ & $  24.710$  \\ 
 8039225 & $ 1.79413$ & $ 7.57$ & $ 6690$ & $ 0.578$ & $ 1.41$ & $3.591$ & $  43.788 \pm   18.072$ & $   4.421$ & $  21.969$  \\ 
 8177958 & $ 1.23526$ & $ 6.40$ & $ 6776$ & $ 1.008$ & $ 1.44$ & $0.593$ & $-160.207 \pm   48.188$ & $ -90.454$ &     ---  \\ 
 8257903 & $ 0.51506$ & $ 3.46$ & $ 6111$ & $ 0.601$ & $ 1.22$ & $1.426$ & $   9.988 \pm    0.265$ & $  18.532$ & $  20.146$  \\ 
 8431389 & $ 0.35109$ & $ 2.73$ & $ 6192$ & $ 0.050$ & $ 1.25$ & $0.760$ & $ -15.623 \pm    0.205$ & $ -58.611$ &     ---  \\ 
 8539720 & $ 0.74450$ & $ 4.33$ & $ 6574$ & $ 0.540$ & $ 1.37$ & $0.428$ & $ -12.018 \pm    0.682$ & $  -5.531$ &     ---  \\ 
 8545456 & $ 0.31520$ & $ 3.07$ & $ 8549$ & $ 0.391$ & $ 1.98$ & $0.970$ & $  -4.635 \pm    0.151$ & $-312.533$ &     ---  \\ 
 8685306 & $ 0.80808$ & $ 5.77$ & $ 9210$ & $ 0.892$ & $ 2.17$ & $1.236$ & $  69.655 \pm    3.404$ & $ 264.794$ & $ 169.576$  \\ 
 8690104 & $ 0.40877$ & $ 2.91$ & $ 6077$ & $ 0.577$ & $ 1.21$ & $1.591$ & $  12.093 \pm    0.252$ & $  20.190$ & $  29.149$  \\ 
 8696274 & $ 0.39149$ & $ 2.75$ & $ 6826$ & $ 1.010$ & $ 1.45$ & $0.250$ & $  -2.422 \pm    0.161$ & $  -1.000$ &     ---  \\ 
 8703528 & $ 0.39987$ & $ 2.93$ & $ 5626$ & $ 0.189$ & $ 1.06$ & $1.025$ & $  16.079 \pm    0.172$ & $ 577.406$ & $  42.084$  \\ 
 8823666 & $ 0.43245$ & $ 3.22$ & $ 6160$ & $ 0.249$ & $ 1.24$ & $1.091$ & $  10.634 \pm    0.295$ & $ 111.450$ & $  29.157$  \\ 
 8872737 & $ 0.45910$ & $ 2.93$ & $ 5410$ & $ 1.022$ & $ 0.99$ & $1.645$ & $  12.360 \pm    0.996$ & $  13.713$ & $  21.333$  \\ 
 9020289 & $ 0.38403$ & $ 2.64$ & $ 6382$ & $ 0.995$ & $ 1.31$ & $3.765$ & $  18.065 \pm    0.560$ & $   7.426$ & $  38.982$  \\ 
 9026766 & $ 0.27213$ & $ 2.16$ & $ 6127$ & $ 0.576$ & $ 1.23$ & $2.095$ & $   6.614 \pm    0.104$ & $   9.075$ & $  22.021$  \\ 
 9110213 & $ 0.33705$ & $ 2.16$ & $ 4891$ & $ 0.159$ & $ 0.80$ & $0.479$ & $ -39.554 \pm    0.183$ & $ -28.990$ &     ---  \\ 
 9145707 & $ 0.32077$ & $ 2.19$ & $ 5557$ & $ 0.934$ & $ 1.04$ & $3.235$ & $   4.035 \pm    0.090$ & $   1.944$ & $   8.530$  \\ 
 9181877 & $ 0.32101$ & $ 2.14$ & $ 4783$ & $ 0.922$ & $ 0.77$ & $1.503$ & $  32.719 \pm    0.278$ & $  51.788$ & $  65.071$  \\ 
 9268105 & $ 0.42569$ & $ 2.82$ & $ 6105$ & $ 0.433$ & $ 1.22$ & $2.821$ & $   4.174 \pm    0.365$ & $   2.188$ & $   8.096$  \\ 
 9283826 & $ 0.35652$ & $ 2.69$ & $ 5785$ & $ 0.282$ & $ 1.11$ & $0.830$ & $ -24.082 \pm    0.465$ & $-122.352$ &     ---  \\ 
 9453192 & $ 0.71884$ & $ 4.49$ & $ 7232$ & $ 0.674$ & $ 1.58$ & $2.084$ & $   3.297 \pm    1.150$ & $   2.225$ & $   5.354$  \\ 
 9840412 & $ 0.87846$ & $ 5.26$ & $ 7018$ & $ 0.292$ & $ 1.51$ & $0.666$ & $-119.139 \pm    1.178$ & $-136.368$ &     ---  \\ 
 9947924 & $ 0.36281$ & $ 2.86$ & $ 6090$ & $ 0.075$ & $ 1.21$ & $0.958$ & $ -23.427 \pm    0.166$ & $-594.043$ &     ---  \\ 
 9956124 & $ 0.36273$ & $ 3.02$ & $ 7033$ & $ 0.960$ & $ 1.52$ & $1.198$ & $  12.806 \pm    1.100$ & $  90.099$ & $  49.096$  \\ 
 9957411 & $ 0.37469$ & $ 2.64$ & $ 6262$ & $ 0.673$ & $ 1.27$ & $2.657$ & $  11.152 \pm    0.529$ & $   7.608$ & $  26.026$  \\ 
10007533 & $ 0.64806$ & $ 3.98$ & $ 7142$ & $ 1.054$ & $ 1.55$ & $3.390$ & $  11.456 \pm    0.768$ & $   3.821$ & $  17.739$  \\ 
10032935 & $ 0.32052$ & $ 2.46$ & $ 5970$ & $ 0.254$ & $ 1.17$ & $0.646$ & $  -7.417 \pm    0.156$ & $ -16.535$ &     ---  \\ 
10084115 & $ 0.30567$ & $ 2.14$ & $ 5859$ & $ 1.070$ & $ 1.14$ & $4.239$ & $  17.279 \pm    0.292$ & $   6.619$ & $  39.745$  \\ 
10154189 & $ 0.41124$ & $ 3.05$ & $ 6202$ & $ 0.444$ & $ 1.25$ & $0.788$ & $  -4.032 \pm    0.274$ & $ -15.199$ &     ---  \\ 
10229723 & $ 0.62872$ & $ 4.14$ & $ 7141$ & $ 0.590$ & $ 1.55$ & $0.542$ & $ -11.897 \pm    0.724$ & $ -11.552$ &     ---  \\ 
10259530 & $ 0.70721$ & $ 4.92$ & $ 7721$ & $ 0.081$ & $ 1.73$ & $1.188$ & $   5.113 \pm    0.717$ & $  22.184$ & $  11.502$  \\ 
10267044 & $ 0.43004$ & $ 3.19$ & $ 6918$ & $ 0.684$ & $ 1.48$ & $1.709$ & $   1.076 \pm    0.164$ & $   1.740$ & $   2.934$  \\ 
10292413 & $ 0.55916$ & $ 3.66$ & $ 7135$ & $ 0.943$ & $ 1.55$ & $0.348$ & $  -2.313 \pm    0.210$ & $  -1.139$ &     ---  \\ 
10322582 & $ 0.29127$ & $ 2.07$ & $ 5844$ & $ 1.094$ & $ 1.13$ & $0.225$ & $ -44.528 \pm    0.381$ & $ -16.766$ &     ---  \\ 
10485137 & $ 0.44527$ & $ 3.18$ & $ 5998$ & $ 0.025$ & $ 1.18$ & $1.203$ & $  15.491 \pm    0.241$ & $  67.515$ & $  37.711$  \\ 
10796477 & $ 0.48497$ & $ 3.69$ & $ 6846$ & $ 0.197$ & $ 1.46$ & $0.963$ & $  -2.548 \pm    0.135$ & $ -67.154$ &     ---  \\ 
11144556 & $ 0.64298$ & $ 4.16$ & $ 6629$ & $ 0.391$ & $ 1.39$ & $1.480$ & $   7.726 \pm    0.296$ & $  11.587$ & $  13.980$  \\ 
11151970 & $ 0.31163$ & $ 2.30$ & $ 5321$ & $ 0.806$ & $ 0.95$ & $0.750$ & $ -29.144 \pm    0.760$ & $ -89.130$ &     ---  \\ 
11305087 & $ 0.30927$ & $ 2.49$ & $ 5732$ & $ 0.179$ & $ 1.10$ & $1.047$ & $  16.855 \pm    0.219$ & $ 423.311$ & $  58.347$  \\ 
11404758 & $ 0.35125$ & $ 2.56$ & $ 6712$ & $ 0.686$ & $ 1.42$ & $0.289$ & $ -28.877 \pm    0.422$ & $ -15.770$ &     ---  \\ 
11494583 & $ 0.24834$ & $ 1.64$ & $ 4713$ & $ 1.044$ & $ 0.74$ & $0.297$ & $ -11.212 \pm    0.947$ & $  -4.707$ &     ---  \\ 
11495781 & $ 0.50793$ & $ 4.24$ & $ 9426$ & $ 1.034$ & $ 2.24$ & $1.326$ & $   3.835 \pm    0.259$ & $  17.292$ & $  14.833$  \\ 
11496078 & $ 0.29972$ & $ 2.54$ & $ 6188$ & $ 0.733$ & $ 1.25$ & $1.055$ & $  11.628 \pm    0.406$ & $ 293.090$ & $  47.099$  \\ 
11509282 & $ 0.63403$ & $ 5.35$ & $11876$ & $ 0.305$ & $ 3.01$ & $1.452$ & $   3.469 \pm    0.598$ & $  12.148$ & $  13.909$  \\ 
11612091 & $ 0.45427$ & $ 3.05$ & $ 7084$ & $ 0.937$ & $ 1.53$ & $5.055$ & $   7.041 \pm    0.188$ & $   1.952$ & $  14.220$  \\ 
11703960 & $ 0.60442$ & $ 3.79$ & $ 6850$ & $ 0.539$ & $ 1.46$ & $2.751$ & $   6.126 \pm    1.050$ & $   2.814$ & $  10.079$  \\ 
11716688 & $ 0.30122$ & $ 2.23$ & $ 5148$ & $ 0.976$ & $ 0.89$ & $1.231$ & $  33.435 \pm    1.095$ & $ 143.340$ & $  90.015$  \\ 
11717798 & $ 0.37471$ & $ 2.67$ & $ 5496$ & $ 0.187$ & $ 1.02$ & $1.276$ & $  38.862 \pm    0.572$ & $ 127.371$ & $  93.906$  \\ 
11910076 & $ 0.34812$ & $ 2.44$ & $ 5601$ & $ 0.994$ & $ 1.05$ & $0.530$ & $ -65.264 \pm    1.074$ & $ -73.953$ &     ---  \\ 
11924311 & $ 0.44512$ & $ 3.38$ & $ 6877$ & $ 0.045$ & $ 1.47$ & $1.307$ & $   6.822 \pm    0.162$ & $  24.401$ & $  19.845$  \\ 
12016304 & $ 1.02405$ & $ 5.33$ & $ 6884$ & $ 0.780$ & $ 1.47$ & $3.208$ & $  17.576 \pm    3.014$ & $   3.807$ & $  16.540$  \\ 
12055014 & $ 0.49990$ & $ 3.44$ & $ 6896$ & $ 0.840$ & $ 1.47$ & $0.473$ & $  -6.440 \pm    0.219$ & $  -5.668$ &     ---  \\ 
12267718 & $ 0.54505$ & $ 3.61$ & $ 6983$ & $ 0.931$ & $ 1.50$ & $2.445$ & $   2.498 \pm    0.295$ & $   1.586$ & $   4.844$  \\ 
\end{longtable}

\begin{ack}
Some of the data presented in this paper were obtained from the Mikulski Archive for Space Telescopes (MAST). 
STScI is operated by the Association of Universities for Research in Astronomy, Inc., under NASA contract NAS5-26555. 
Support for MAST for non-HST data is provided by the NASA Office of Space Science via grant NNX09AF08G and by other grants and contracts.
This paper includes data collected by the Kepler Mission. Funding for the Kepler Mission is provided by the NASA Science Mission directorate.
The author is grateful to the Chukyo University Research Fund for financial assistance with this research.
We are grateful to the referee, Dr Kazimierz {St{\c e}pie{\'n}}, for suggesting many valuable comments that improved this paper. 
\end{ack}


\begin{thebibliography}{35}
\expandafter\ifx\csname natexlab\endcsname\relax\def\natexlab#1{#1}\fi

\bibitem[{{Applegate}(1992)}]{Applegate1992-ApJ}
{Applegate}, J.~H. 1992, \apj, 385, 621

\bibitem[{{Armstrong} {et~al.}(2014){Armstrong}, {G{\'o}mez Maqueo Chew},
  {Faedi}, \& {Pollacco}}]{Armstrong2014-MNRAS}
{Armstrong}, D.~J., {G{\'o}mez Maqueo Chew}, Y., {Faedi}, F., \& {Pollacco}, D.
  2014, \mnras, 437, 3473

\bibitem[{{Binnendijk}(1970)}]{Binnendijk1970-VA}
{Binnendijk}, L. 1970, Vistas in Astronomy, 12, 217

\bibitem[{{Borkovits} \& {Hegedues}(1996)}]{Borkovits1996-AAS}
{Borkovits}, T. \& {Hegedues}, T. 1996, \aaps, 120, 63

\bibitem[{{Deb} \& {Singh}(2011)}]{Deb2011-MNRAS}
{Deb}, S. \& {Singh}, H.~P. 2011, \mnras, 412, 1787

\bibitem[{{Eker} {et~al.}(2006){Eker}, {Demircan}, {Bilir}, \& {Karata{\c
  s}}}]{Eker2006-MNRAS}
{Eker}, Z., {Demircan}, O., {Bilir}, S., \& {Karata{\c s}}, Y. 2006, \mnras,
  373, 1483

\bibitem[{{Flannery}(1976)}]{Flannery1976-ApJ}
{Flannery}, B.~P. 1976, \apj, 205, 217

\bibitem[{{Frieboes-Conde} \& {Herczeg}(1973)}]{Frieboes1973-AAS}
{Frieboes-Conde}, H. \& {Herczeg}, T. 1973, \aaps, 12, 1

\bibitem[{{Gazeas} \& {St{\c e}pie{\'n}}(2008)}]{Gazeas2008-MNRAS}
{Gazeas}, K. \& {St{\c e}pie{\'n}}, K. 2008, \mnras, 390, 1577

\bibitem[{{Gazeas} \& {Niarchos}(2006)}]{Gazeas2006-MNRAS}
{Gazeas}, K.~D. \& {Niarchos}, P.~G. 2006, \mnras, 370, L29

\bibitem[{{Gilliland} {et~al.}(2010){Gilliland}, {Jenkins}, {Borucki},
  {Bryson}, {Caldwell}, {Clarke}, {Dotson}, {Haas}, {Hall}, {Klaus}, {Koch},
  {McCauliff}, {Quintana}, {Twicken}, \& {van Cleve}}]{Gilliland2010-ApJ}
{Gilliland}, R.~L., {Jenkins}, J.~M., {Borucki}, W.~J., {et~al.} 2010, \apjl,
  713, L160

\bibitem[{{Harmanec}(1988)}]{Harmanec1988-BAICz}
{Harmanec}, P. 1988, Bulletin of the Astronomical Institutes of Czechoslovakia,
  39, 329

\bibitem[{{Hilditch}(2001)}]{Hilditch2001-icbs}
{Hilditch}, R.~W. 2001, {An Introduction to Close Binary Stars}, 392

\bibitem[{{Irwin}(1959)}]{Irwin1959-AJ}
{Irwin}, J.~B. 1959, \aj, 64, 149

\bibitem[{{Jenkins} {et~al.}(2010){Jenkins}, {Caldwell}, {Chandrasekaran},
  {Twicken}, {Bryson}, {Quintana}, {Clarke}, {Li}, {Allen}, {Tenenbaum}, {Wu},
  {Klaus}, {Van Cleve}, {Dotson}, {Haas}, {Gilliland}, {Koch}, \&
  {Borucki}}]{Jenkins2010-ApJ}
{Jenkins}, J.~M., {Caldwell}, D.~A., {Chandrasekaran}, H., {et~al.} 2010,
  \apjl, 713, L120

\bibitem[{{Koch} {et~al.}(2010){Koch}, {Borucki}, {Basri}, {Batalha}, {Brown},
  {Caldwell}, {Christensen-Dalsgaard}, {Cochran}, {DeVore}, {Dunham},
  {Gautier}, {Geary}, {Gilliland}, {Gould}, {Jenkins}, {Kondo}, {Latham},
  {Lissauer}, {Marcy}, {Monet}, {Sasselov}, {Boss}, {Brownlee}, {Caldwell},
  {Dupree}, {Howell}, {Kjeldsen}, {Meibom}, {Morrison}, {Owen}, {Reitsema},
  {Tarter}, {Bryson}, {Dotson}, {Gazis}, {Haas}, {Kolodziejczak}, {Rowe}, {Van
  Cleve}, {Allen}, {Chandrasekaran}, {Clarke}, {Li}, {Quintana}, {Tenenbaum},
  {Twicken}, \& {Wu}}]{Koch2010-ApJ}
{Koch}, D.~G., {Borucki}, W.~J., {Basri}, G., {et~al.} 2010, \apjl, 713, L79

\bibitem[{{Lucy}(1976)}]{Lucy1976-ApJ}
{Lucy}, L.~B. 1976, \apj, 205, 208

\bibitem[{{Paczy{\'n}ski} {et~al.}(2006){Paczy{\'n}ski}, {Szczygie{\l}},
  {Pilecki}, \& {Pojma{\'n}ski}}]{Paczynski2006-MNRAS}
{Paczy{\'n}ski}, B., {Szczygie{\l}}, D.~M., {Pilecki}, B., \& {Pojma{\'n}ski},
  G. 2006, \mnras, 368, 1311

\bibitem[{{Pr{\v s}a} {et~al.}(2011){Pr{\v s}a}, {Batalha}, {Slawson}, {Doyle},
  {Welsh}, {Orosz}, {Seager}, {Rucker}, {Mjaseth}, {Engle}, {Conroy},
  {Jenkins}, {Caldwell}, {Koch}, \& {Borucki}}]{Prsa2011-AJ}
{Pr{\v s}a}, A., {Batalha}, N., {Slawson}, R.~W., {et~al.} 2011, \aj, 141, 83

\bibitem[{{Qian}(2001)}]{Qian2001-MNRAS}
{Qian}, S. 2001, \mnras, 328, 635

\bibitem[{{Qian}(2002)}]{Qian2002-MNRAS}
{Qian}, S. 2002, \mnras, 336, 1247

\bibitem[{{Qian}(2003)}]{Qian2003-MNRAS}
{Qian}, S. 2003, \mnras, 342, 1260

\bibitem[{{Robertson} \& {Eggleton}(1977)}]{Robertson1977-MNRAS}
{Robertson}, J.~A. \& {Eggleton}, P.~P. 1977, \mnras, 179, 359

\bibitem[{{Rucinski}(2007)}]{Rucinski2007-MNRAS}
{Rucinski}, S.~M. 2007, \mnras, 382, 393

\bibitem[{{Rucinski} {et~al.}(2007){Rucinski}, {Pribulla}, \& {van
  Kerkwijk}}]{Rucinski2007-AJ}
{Rucinski}, S.~M., {Pribulla}, T., \& {van Kerkwijk}, M.~H. 2007, \aj, 134,
  2353

\bibitem[{{Sarna}(1993)}]{Sarna1993-MNRAS}
{Sarna}, M.~J. 1993, \mnras, 262, 534

\bibitem[{{Slawson} {et~al.}(2011){Slawson}, {Pr{\v s}a}, {Welsh}, {Orosz},
  {Rucker}, {Batalha}, {Doyle}, {Engle}, {Conroy}, {Coughlin}, {Gregg},
  {Fetherolf}, {Short}, {Windmiller}, {Fabrycky}, {Howell}, {Jenkins}, {Uddin},
  {Mullally}, {Seader}, {Thompson}, {Sanderfer}, {Borucki}, \&
  {Koch}}]{Slawson2011-AJ}
{Slawson}, R.~W., {Pr{\v s}a}, A., {Welsh}, W.~F., {et~al.} 2011, \aj, 142, 160

\bibitem[{{St{\c e}pie{\'n}}(2006{\natexlab{a}})}]{Stepien2006-AcA199}
{St{\c e}pie{\'n}}, K. 2006{\natexlab{a}}, \actaa, 56, 199

\bibitem[{{St{\c e}pie{\'n}}(2006{\natexlab{b}})}]{Stepien2006-AcA347}
{St{\c e}pie{\'n}}, K. 2006{\natexlab{b}}, \actaa, 56, 347

\bibitem[{{Tokovinin} {et~al.}(2006){Tokovinin}, {Thomas}, {Sterzik}, \&
  {Udry}}]{Tokovinin2006-AA}
{Tokovinin}, A., {Thomas}, S., {Sterzik}, M., \& {Udry}, S. 2006, \aap, 450,
  681

\bibitem[{{Vilhu}(1982)}]{Vilhu1982-AA}
{Vilhu}, O. 1982, \aap, 109, 17

\bibitem[{{Webbink}(2003)}]{Webbink2003-ASPC}
{Webbink}, R.~F. 2003, in Astronomical Society of the Pacific Conference
  Series, Vol. 293, 3D Stellar Evolution, ed. S.~{Turcotte}, S.~C. {Keller}, \&
  R.~M. {Cavallo}, 76

\bibitem[{{Yang} {et~al.}(2011){Yang}, {Shao}, {Pan}, \& {Yin}}]{Yang2011-PASP}
{Yang}, Y.-G., {Shao}, Z.-Y., {Pan}, H.-J., \& {Yin}, X.-G. 2011, \pasp, 123,
  895

\bibitem[{{Yang} \& {Wei}(2009)}]{Yang2009-AJ137}
{Yang}, Y.-G. \& {Wei}, J.-Y. 2009, \aj, 137, 226

\bibitem[{{Zhu} {et~al.}(2009){Zhu}, {Qian}, {Zola}, \& {Kreiner}}]{Zhu2009-AJ}
{Zhu}, L.~Y., {Qian}, S.~B., {Zola}, S., \& {Kreiner}, J.~M. 2009, \aj, 137,
  3574

\end{thebibliography}
\end{document}